%% file: shortcohqph.tex
\documentclass[aps,prl,twocolumn,groupedaddress,amssymb,amsfonts,showpacs]{revtex4}

\usepackage{color}
\usepackage[dvips]{graphics,graphicx}
\input{ekqc_a}

\newcommand{\shortqph}[1]{#1}

\providecommand{\ignore}[1]{}

\def\R{{\mathbb{R}}}

\def\openone{\leavevmode\hbox{\small1\kern-3.8pt\normalsize1}}
\def\RR{{\rm I\kern-.2emR}}
\def\tr{{\rm tr}\; }
\def\fu{\mathfrak{u}}

\def\fg{\mathfrak{g}}
\def\fh{\mathfrak{h}}

\def\fsu{\mathfrak{su}}

\def\fso{\mathfrak{so}}

\def\openone{\leavevmode\hbox{\small1\kern-3.8pt\normalsize1}}
\def\RR{{\rm I\kern-.2emR}}
\def\tr{{\rm tr}\; }

\def\ch{{\cal H}}

\providecommand{\ignore}[1]{}

\renewcommand{\ket}[1]{| #1 \rangle}
\renewcommand{\bra}[1]{\langle #1 |}
\newcommand{\proj}[1]{\ket{#1}\! \bra{#1}}

\newcommand{\melement}[2]{ \langle #1 | #2 | #1 \rangle}
\newcommand{\bitem}{\begin{itemize}}
\newcommand{\eitem}{\end{itemize}}
\newcommand{\benum}{\begin{enumerate}}
\newcommand{\eenum}{\end{enumerate}}
\newcommand{\beq}{\begin{equation}}
\newcommand{\eeq}{\end{equation}}
\newcommand{\beqa}{\begin{eqnarray}}
\newcommand{\eeqa}{\end{eqnarray}}
\newtheorem{definition}{Definition}

\newtheorem{proposition}{Proposition}

\newcommand{\bproof}{\begin{proof}}
\newcommand{\eproof}{\end{proof}}
\newcommand{\bprop}{\begin{proposition}}

\newcommand{\bdef}{\begin{definition}}
\newcommand{\zls}{{\mathbb Z}}

\begin{document}


\title{A subsystem-independent generalization of entanglement}



\author{Howard Barnum}
\shortqph{\email[]{barnum@lanl.gov}}
\author{Emanuel Knill}
\shortqph{\email[]{knill@lanl.gov}}
\author{Gerardo Ortiz}
\shortqph{\email[]{g_ortiz@lanl.gov}}
\author{Rolando Somma}
\shortqph{\email[]{somma@lanl.gov}}
\author{Lorenza Viola}
\shortqph{\email[]{lviola@lanl.gov}}
\affiliation{Los Alamos National Laboratory, MS B256, Los Alamos, NM 87545}


\date{\today}

\begin{abstract}
We introduce a generalization of entanglement based on the idea that
entanglement is relative to a distinguished subspace of observables
rather than a distinguished subsystem decomposition. A pure quantum
state is entangled relative to such a subspace if its expectations are
a proper mixture of those of other states. Many information-theoretic
aspects of entanglement can be extended to the general setting,
suggesting new ways of measuring and classifying entanglement 
in multipartite systems. By going beyond the distinguishable-subsystem 
framework, generalized entanglement also provides novel tools for probing 
quantum correlations in interacting many-body systems.
\end{abstract}



\maketitle

Entanglement is a uniquely quantum phenomenon whereby a pure state of
a composite quantum system may cease to be determined by the states of
its constituent subsystems~\cite{Schroedinger}. Entangled pure states
are those that have {\it mixed} subsystem states. To determine an
entangled state requires knowledge of the correlations between the
subsystems. As no pure state of a classical system can be correlated,
such correlations are intrinsically non-classical, as strikingly
manifested by the violation of local realism and Bell's
inequalities~\cite{Bell93a}. In the science of quantum information
processing (QIP), entanglement is regarded as the defining resource
for quantum communication and an essential feature needed for unlocking the 
power of quantum computation. However, in spite of intensive investigation, 
a complete understanding of entanglement is far from being reached.

To unambiguously define entanglement requires a preferred partition of
the overall system into subsystems. In conventional QIP scenarios,
subsystems are associated with spatially separated ``local'' parties,
which legitimates the \emph{distinguishability} assumption implicit in
standard entanglement theory.  However, because quantum correlations
are at the heart of many physical phenomena, it would be desirable for
a notion of entanglement to be useful in contexts other than QIP.
Strongly interacting quantum systems offer compelling examples of
situations where the usual subsystem-based view is inadequate.
Whenever indistinguishable particles are sufficiently close to each
other, quantum statistics forces the accessible state space to be a
proper subspace of the full tensor product space, and exchange
correlations arise that are not a usable resource in the usual QIP
sense. Thus, the natural identification of particles with preferred
subsystems becomes problematic.  Even if a distinguishable-subsystem
structure may be associated to degrees of freedom different from the
original particles (such as a set of modes~\cite{Zanardi2001a}),
inequivalent factorizations may occur on the same footing. Finally,
the introduction of quasiparticles, or the purposeful transformation
of the algebraic language used to analyze the system~\cite{Batista},
may further complicate the choice of preferred subsystems. While
efforts are under way to obtain entanglement-like notions for bosons
and fermions~\cite{Eckert2002a,Zanardi2001a} and to study entanglement
in quantum critical phenomena~\cite{Osborne,Osterloh2002a,Vidal2002a},
formulating a theory of entanglement applicable to the full variety of
physical settings remains an important challenge.

In this Letter, we introduce a notion of \emph{generalized
entanglement} (GE) based on the relationship of a state to different
sets of observables of the system of interest, without reference to a
preferred subsystem decomposition. This is achieved by realizing that
the salient features of entanglement are determined by the
expectations of a \emph{distinguished subspace of observables}. The
latter may represent a limited means of manipulating and observing the
system. For standard entanglement these means are limited to local
observables acting on one subsystem only. The central idea is to
generalize the observation that standard entangled pure states are
those that look mixed to local observers. Each pure quantum state
gives rise to a \emph{reduced} state that only provides the
expectations of the distinguished observables. The set of reduced
states is convex and, like an ordinary quantum state space, it
includes pure states (the extremal ones). We say that a pure state is
\emph{generalized unentangled relative to the distinguished
observables}, if its reduced state is pure, and \emph{generalized
entangled} otherwise.  The definition extends to mixed states in a
standard way: A mixed state is unentangled if it can be written as a
mixture (or convex combination) of unentangled pure states.  Because
our definition depends only on convex properties of the distinguished
spaces of observables and states we consider, it provides a notion of
entanglement within a general convex framework suitable for investigating 
the foundations of quantum mechanics and related physical theories 
(cfr.~\cite{Beltrametti97a} and references therein).

The mathematical foundation of GE is established in~\cite{BKOV2002a}.
Here we highlight the significance of GE from a physics and
information-physics perspective. For this purpose, we focus on the
case where the observable subspace is a Lie algebra. A key result is
then the identification of pure generalized unentangled states with
the \emph{generalized coherent states} (GCSs, a connection
independently noted by Klyachko~\cite{klyachko2002a}), which are well
known for their applications in physics~\cite{Zhang90a}.  This
encompasses the entanglement settings introduced to date in a unifying
framework.  Furthermore, it is now possible to extend
information-theoretic notions to coherent state theory and beyond. We
demonstrate that many concepts previously thought to be
subsystem-specific are much more generic, define new measures of
entanglement based on the general theory, and apply quantum
information to condensed-matter problems. In particular, we introduce
notions of \emph{Generalized Local Operations assisted by Classical
Communication} (GLOCC) under which the ordinary measures of standard
entanglement do not increase, as well as measures of GE with the
desired behavior under classes of GLOCC maps.  New measures of
standard entanglement are obtained for the multipartite case. In the
Lie-algebraic setting, a simple GE measure obtained from the
\emph{purity relative to a Lie algebra} is a useful diagnostic tool
for quantum many-body systems, playing the role of a \emph{disorder
parameter} for broken-symmetry quantum phase transitions.

\textbf{Generalized entanglement.$-$} We first revisit the standard
setting for entanglement where we have two distinguishable subsystems
forming a bipartite system. Let the $mn$-dimensional joint state space
$\ch$ factorize as $\ch= \ch_a \otimes \ch_b$, with $\ch_a$, $\ch_b$
$m$, $n$-dimensional, respectively. In this setting, physical
considerations distinguish a preferred set of observables, spanned by
traceless Hermitian operators of the form $A \otimes \openone$ and
$\openone \otimes B$, which are the local observables acting on system
$a$ or $b$ alone. For each pure state $|\psi\rangle \in \ch_a
\otimes \ch_b$, one may consider the reduced state describing the
expectations of measurements of local observables.  The reduced
state is determined by the pair of reduced density operators, $\rho_a
:= \text{tr}_b |\psi\rangle\langle\psi|$ and $\rho_b := \text{tr}_a
|\psi\rangle\langle\psi|$. Because pure product states are exactly
those for which subsystem states are pure, our definition of GE
\emph{relative to the local observable subspace} coincides with the
standard definition of entanglement.
\shortqph{%
In this example,
the distinguished observable space is a Lie algebra, $\fh= \fsu(m)
\oplus \fsu(n)$, and $\fh$ is a subalgebra of the full Lie algebra
$\fg$ of operators on $\ch$.  The connection with GCSs is established
by associating the family of pure unentangled states with an orbit of
the group of local unitary transformations acting on $\ch$ (see
below).
}%

The extent to which our viewpoint extends the usual subsystem-based
definition may be appreciated in situations where no subsystem
partition exists and conventional entanglement is meaningless.
Consider a single spin-1 system, whose three-dimensional state space
$\ch$ carries an irreducible representation of $\fsu(2)$, with
generators $J_x, J_y, J_z$ satisfying $[J_\alpha, J_\beta] = i
\varepsilon_{\alpha\beta\gamma} J_\gamma$,
($\varepsilon_{\alpha\beta\gamma}$ being the totally antisymmetric
tensor). Suppose that the distinguished observables are linear in
these generators so that they are the ones in the given representation
of $\fsu(2)$. The reduced states can be identified with vectors of
expectation values of these three observables: They form a unit ball
in $\R^3$, and the extremal points are those on the surface, 
which have maximal spin component $1$ for some linear combination
of $J_x, J_y, J_z$.  These are the well-known ``spin coherent
states,'' or GCSs for SU(2)~\cite{Zhang90a}. For any choice of spin
direction, $\ch$ is spanned by the $|1\rangle, |0\rangle,
|{-1}\rangle$ eigenstates of that spin component; the first and last
are GCSs, but $|0\rangle$ is \emph{not}, characterizing $|0\rangle$ as
a generalized entangled state relative to $\fsu(2)$. All pure states
appear unentangled if access to the full algebra $\fg=\fsu(3)$ is
available (that is, $\fsu(3)$ is distinguished).
   
This example illustrates that when the distinguished subspace 
forms an irreducibly represented Lie algebra, the set of 
unentangled states is the set of GCSs. Another,
more physically motivated characterization is as the set of states
that are unique ground states of a distinguished observable. To
formally relate these characterizations of unentangled states we 
review the needed Lie representation theory~\cite{Humphreys72a}.  A
\emph{Cartan subalgebra} (CSA) $\lie{c}$ of a semisimple Lie algebra 
$\lie{h}$ is a maximal commutative subalgebra. A vector space carrying 
a representation of $\lie{h}$ decomposes into orthogonal joint 
eigenspaces $V_\lambda$ of the operators in $\lie{c}$. That is, each
$V_\lambda$ consists of the set of states $\ket{\psi}$ such  that for
$x\in\lie{c}$, $x\ket{\psi}=\lambda(x)\ket{\psi}$. The label $\lambda$
is therefore a linear functional on $\lie{c}$, called the
\emph{weight} of $V_\lambda$. In the above example, any spin
component $J_\alpha$ spans a (one-dimensional) CSA $\lie{c}_\alpha$. 
There are three  weight spaces labeled by the angular momentum along
$\alpha$, and spanned by the states $|{1}\rangle, |{0}\rangle,
|{-1}\rangle$ of the  previous paragraph.  
\shortqph{%
Note that any two CSAs are conjugate under elements of the Lie
group, manifested in the spin example by the fact that $J_\alpha$
transforms into any desired spin component via conjugation by a
rotation in SU(2). 
}%
The subspace of operators in $\lie{h}$ orthogonal in the trace inner 
product to $\lie{c}$ can be organized into orthogonal ``raising and
lowering'' operators, which connect different weight spaces. In the
example, choosing $J_z$ as the basis of our CSA, these are $J_{\pm} :=
(J_x \pm iJ_y)/\sqrt{2}$. For a fixed CSA and irreducible
representation, the  weights generate a convex
polytope; a lowest (or highest) weight is an extremal  point of such a
polytope, and the one-dimensional weight-spaces  having those weights
are known as \emph{lowest-weight states}.  The set of lowest-weight
states for all CSAs is the orbit of any one such state under the Lie
group generated by $\lie{h}$.  These are the group-theoretic GCSs
\cite{Zhang90a}. Notably, the GCSs attain \emph{minimum uncertainty}
in an appropriate invariant sense~\cite{Delbourgo77b}.
  
A natural way to relate any state $|\psi\rangle \in \ch$ to a Lie
algebra $\fh$ of operators acting on $\ch$ is to project $\proj{\psi}$
onto $\fh$.  This projection completely determines the expectations of
operators in $\lie{h}$ for $\ket{\psi}$. The generalized unentangled
states are the ones for which the projection of $\proj{\psi}$ onto
$\fh$ is extremal. \shortqph{The intuition that these should be the
states whose projection has largest distance from $0$ turns out to be
true.} This motivates the following definition.  Let $\{x_i\}$ be a
Hermitian ($x_i=x_i^\dagger$) orthogonal ($\tr{x_i x_j} \propto
\delta_{ij}$) basis for $\fh$~\cite{comment1}. The \emph{purity of
$\ket{\psi}$ relative to $\fh$} (or \emph{$\fh$-purity}) is
$P_\fh (|\psi\rangle) :=
\sum_i |\melement{\psi}{x_i}|^2$,
where the $x_i$ have a common, rescaled norm chosen to ensure that
the maximal value is 1. $P_\fh (|\psi\rangle)$ is the square-distance
from $0$ of the projection of $\proj{\psi}$. For pure bipartite
states, the $ \fsu(m) \oplus \fsu(n)$-purity is (up to a constant) the 
conventional purity given by the trace of the square of either subsystem's 
reduced density operator.

So far, $\fh$ has been assumed to be a \emph{real} Lie algebra of
Hermitian operators.  These may be thought of as a preferred family of
Hamiltonians, which generate (via $h \mapsto e^{ih}$) a Lie group of
unitary operators.  More generally, we want Lie-algebraically
distinguished completely positive (CP) maps, $\rho \mapsto \sum_i A_i
\rho A_i^\dagger$.  A natural class is obtained by restricting the
``Hellwig-Kraus'' (HK) operators $A_i$ to lie in the topological
closure $\overline{e^{\fh_c\oplus \one}}$ of the Lie group generated
by the \emph{complex} Lie algebra $\fh_c\oplus
\one$~\cite{comment2}. Having HK operators in a group ensures closure
under composition.  Using $\fh_c \oplus \one$ allows non-unitary HK
operators.  Topological closure introduces singular operators such as
projectors.  The following characterizations of unentangled
states (proven in~\cite{BKOV2002a}) demonstrate 
the power of the Lie algebraic setting.

\par

\textbf{Theorem.}
\newcounter{stcharcnt}
The following are equivalent for an irreducible representation of 
$\fh$ on $\ch$:
\\  \noindent \label{st1}
  {\textbf{(1)}} $\rho$ is generalized unentangled relative to $\lie{h}$.
\\ \noindent \label{st2}
  {\textbf{(2)}} $\rho=|\psi\rangle\langle\psi|$ with 
  $|\psi\rangle$ the unique ground state of 
  some \\ \hspace*{6mm} $H$ in $\lie{h}$.
\\ \noindent \label{st3}
  {\textbf{(3)}} $\rho=|\psi\rangle\langle\psi|$ with
  $|\psi\rangle$ a lowest-weight vector 
of $\lie{h}$.
\\ \noindent \label{st5}
  {\textbf{(4)}} $\rho$ has maximum $\lie{h}$-purity.
\\ \noindent \label{st6}
  {\textbf{(5)}} $\rho$ is a one-dimensional projector in
  $\overline{e^{\lie{\fh_c \oplus \one}}}$.

\par

\textbf{Generalized LOCC.$-$} The semigroup of LOCC
maps~\cite{Bennett96c} and the preordering it induces on states
\shortqph{according to whether or not a given state can be transformed
to another by an LOCC operation} are at the core of entanglement
theory.  Given an HK representation $\{A_i\}$ of a CP map $M$, we can
view each $A_i$ as being associated with measurement outcome $i$,
obtained with probability $\tr A_i \rho$, and leading to the state
$A_i \rho A_i^\dagger$.  The set $\{ A_i\}$ and a list of maps $M_i$,
with HK operators $\{B_{ij}\}$, specify a new map with representation
$\{B_{ij}A_i\}$. This map can be implemented by first applying $M$ and
then, given measurement outcome $i$, applying $M_i$. We call this {\em
conditional composition} of maps.  Closing the set of one-party maps
(for all parties) under conditional composition gives the LOCC
maps. When the distinguished observables form a semisimple Lie algebra
$\fh$, a natural multipartite structure can be exploited to generalize
LOCC. $\fh$ can be uniquely expressed as a direct sum of simple Lie
algebras, $\fh = \oplus_i \fh_i$. A Hilbert space irreducibly
representing $\fh$ factorizes as $\ch = \otimes_i \ch_i$, with $\fh_i$
acting non-trivially on $\ch_i$ only.  This resembles ordinary
entanglement, except that the ``local'' systems $\ch_i$ may not be
\emph{physically} local, and actions on them are restricted to involve
operators in the topological closure
of a ``local'' Lie group representation which need not be
GL(\text{dim}$(\ch_i))$ as in standard entanglement.  For each simple
algebra $\fh_i$, a natural restriction is to CP maps with HK operators
in $\overline{e^{(\fh_i)_c \oplus \one}}$.  GLOCC, generalized LOCC,
is the closure under conditional composition of the set of operations
each of which is representable with HK operators in the topological
closure of $e^{(\fh_i)_c \oplus \one}$ for some $i$.  

\shortqph{ In conventional entanglement, there is also interest in
\emph{separable} maps (SLOCC,
~\cite{Vidal2000b,Bennett2001a,Duer2001a}), which are those
representable with HK operators that are tensor products. The
generalization of these maps is obtained by considering the semigroup
of maps whose HK operators are in $\overline{e^{\fh_c\oplus\one}}$.
Another potential generalization of LOCC involves using spectra of
operators to classify them as analogues of {\it single-party}
operators.  Yet another begins from maps that induce well-defined maps
on the set of reduced states, as single-party maps do in the standard
setting.  These alternative proposals are discussed further
in~\cite{BKOV2002a}.  }%

\textbf{Measures of generalized entanglement.$-$} Because GE relative
to $\fh$ reflects \emph{incoherence} relative to $\fh$, and
incoherence amounts to \emph{mixing} from the point of view of $\fh$,
a natural Lie-algebraic entanglement measure for mixed $\rho$ is
obtained by minimizing the expected difference from $1$ of the
$\fh$-purity over pure state ensembles for $\rho$: $\min\left(1 -
\sum_i p_i P_\fh(\pi_i)\right)$, where the minimum is over all $p_i>0$
and pure $\pi_i$ such that $\sum_i p_i \pi_i = \rho$.  A different
approach is suggested by the convex structure of reduced states and
uses natural \emph{mixedness measures} $\sigma$ on finite probability
distributions $\mathbf{p}=(p_1,\ldots, p_k)$. Such measures are
concave and permutation-invariant\shortqph{(\emph{Schur concave})}.
Examples are entropy, $\sigma_{\ln}(\mathbf{p}) := -\sum_i p_i \ln
p_i$, and Renyi entropy, $\sigma_1(\mathbf{p}):= 1-\sum_i p_i^2$.  For
a reduced state $\mu$, define $\sigma(\mu)$ by minimizing $\sigma(p)$
over ways of writing $\mu=\sum_i p_i \mu_i$ with $p_i$ probabilities
and $\mu_i$ pure reduced states.  For an unreduced pure state $\rho$
with reduction $\nu$, define $\sigma(\rho):= \sigma(\nu)$.  For
general unreduced $\rho$, define $\sigma(\rho) = \min \sum_i p_i
\sigma(\pi_i)$ where the minimum is over all $p_i>0$ and pure $\pi_i$
such that $\sum_i p_i \pi_i = \rho$.  This measure will be convex as
all measures of GE should be.  It is also desirable that it is
non-increasing under GLOCC.  In~\cite{BKOV2002a}, we have established
that the above measures are non-increasing under those GLOCC
operations implementable via conditional composition of operations
with \emph{unitary} HK operators in the Lie group.
\shortqph{Generalizations of these results beyond the Lie-algebraic
setting are discussed in~\cite{BKOV2002a}.} As with standard
entanglement, no single measure can capture the complexity of GE.

\textbf{Generalized multipartite entanglement.$-$} Multipartite
systems are examples where GE contributes to the study of conventional
entanglement.  For $N$ qubits, the relevant algebra for conventional
entanglement is $\fh= \oplus_{i =1}^{N} \fsu(2)_i$, generated by the
Pauli matrices for each qubit. The pure product states have maximal
purity $P_\fh=1$ (unentangled), whereas the states $|{\rm GHZ}_N
\rangle := 2^{-1/2} [ |{\uparrow \uparrow \cdots \uparrow}\rangle +
|{\downarrow \downarrow \cdots \downarrow} \rangle]$ have minimal
purity $0$ (maximally entangled).  States of the form $|{\rm
W}_N\rangle := N^{-1/2} \sum_{i=1}^N |{\uparrow \uparrow \cdots
\uparrow \downarrow_i \uparrow \cdots \uparrow}\rangle $ have an
intermediate purity $(\frac{N-2}{N})^2$. In the $N\rightarrow \infty$
limit, $P_\fh(|{\rm W}_N\rangle)$ $\rightarrow 1$, whereas $|{\rm
GHZ}_N \rangle$ remains maximally entangled.  Interestingly, $1-P_\fh$
(for this $\fh$) coincides with the global entanglement measure
introduced in~\cite{Meyer2002}.  \shortqph{ Different choices of
observable algebras, constructed for instance by using the full
$\fsu(2^k)$ or collective $\fsu(2)$ subalgebras for clusters of $k$
qubits, may be considered to further refine the study of GE. Algebras
such as these can be seen to be partially ordered by the subalgebra
relationship. In this case, one can further relativize entanglement by
considering any pair of relevant Lie algebras
$\lie{h}_1\supseteq\lie{h}_2$. By starting with a state reduced to
$\lie{h}_1$ and considering its further reduction to $\lie{h}_2$,
contributions to entanglement due to this pair of subalgebras can be
extracted. } The measures of entanglement $\sigma(\rho)$ can provide
additional information on the fine-structure of multipartite quantum
correlations.

Another example consists of two spin-1 particles in 
the \emph{total spin} representation of $\fsu(2)$. Suppose that the
two spins can only be accessed collectively, {\it e.g.} using a global
external field. Then the distinguished observable subspace is spanned
by operators $J_\alpha := J^{(1)}_\alpha \otimes \openone + \openone
\otimes J^{(2)}_\alpha$, $J^{(1)}_\alpha, J^{(2)}_\alpha$ being spin-1
generators for each $\fsu(2)$. The (unentangled) GCSs here are states
of maximal total spin projection in some direction $\alpha$ (states of
the form $|{1_\alpha} \rangle |{1_\alpha}\rangle$), whereas product
states, like $|{0}_\alpha\rangle |{0}_\alpha\rangle$ with zero spin
projection, are generalized (maximally) entangled relative to this
algebra. This reflects the fact that no  SU(2)
spin rotation can connect $|{0}_\alpha\rangle |{0}_\alpha\rangle$ to
the unentangled state $|{1_\alpha} \rangle |{1_\alpha}\rangle$.

\textbf{Entanglement in condensed matter.$-$}
GE can be applied to the study of interacting quantum systems,
where the characterization of quantum correlations is essential to a
complete understanding of quantum phase transitions. Consider the case
of an anisotropic one-dimensional spin-1/2  XY model in a transverse
field, described by the Hamiltonian acting on the $N$-spin space:
\begin{equation} \label{Ham1} 
H =-g \sum\limits_{i=1}^{N}  [(1+\eta) J_x^i J_x^{i+1}+ (1-\eta)
J_y^i J_y^{i+1}] + \sum\limits_{i=1}^{N} J_z^i \:,
\end{equation}
where $\eta \in [0,1]$ is the anisotropy, $g \in [0,\infty)$
is a tunable parameter and $J_\alpha^{N+1}=J_\alpha^{1}$. $H$ can be
diagonalized by performing a Jordan-Wigner mapping to spinless
fermions. The resulting ground state is BCS-like. A transition
between a paramagnetic state (disorder) and a ferromagnetic state
(order) occurs for all $\eta$ at the critical value $g_c=1$, in the
thermodynamic limit. Relevant algebras, generated by subsets of
bilinear products of spinless-fermion operators~\cite{Zhang90a},
include $\fu(N) = \{ c^\dagger_i c^{\;}_i -{1\over 2},
\frac{c^\dagger_i c^{\;}_j+c^\dagger_j c^{\;}_i}{\sqrt{2}},
\frac{c^\dagger_i c^{\;}_j-c^\dagger_j c^{\;}_i}{i\sqrt{2}} \}$, and
$\fso(2N)= \fu(N) \oplus \{ \frac{c^\dagger_i c^{\dagger}_j+c^{\;}_j
c^{\;}_i}{\sqrt{2}}, \frac{c^\dagger_i c^{\dagger}_j-c^{\;}_j
c^{\;}_i}{i\sqrt{2}} \}, 1 \le i < j \le N.$ A BCS state is a GCS
of $\fso(2N)$, thus it is generalized-{\it un}entangled relative to
$\fso(2N)$, capturing the fact that quasiparticles are
\emph{non-interacting} in this description. However, GE may be present
relative to the smaller algebra $\fu(N) \subset \fso(2N)$~\cite{comment3}. 
Remarkably, $P_{\fu(N)}$ as a function of $g$ plays the 
role of a disorder parameter (Fig.~1). The fact that the purity relative 
to an appropriate algebra succeeds at detecting a quantum phase transition
and characterizing its universality class appears to be a generic
feature of broken-symmetry (here $\zls_2$) phase transitions. The
purity, a sum of squared expectations of observables, is a
natural measure of \emph{fluctuations}. Changes in the nature of the
fluctuations identify those transitions. In some cases~\cite{Osborne,
Osterloh2002a}, nearest-neighbor lattice-site entanglement or other
standard entanglement measures may suffice, but in general highly
non-local correlations or fluctuations, whose nature depends on the
physics and symmetries of the problem, may be required.  An extended
analysis of these issues will be presented elsewhere~\cite{Somma2003}.

\begin{figure}
\includegraphics[width=3.3in,height=2.1in]{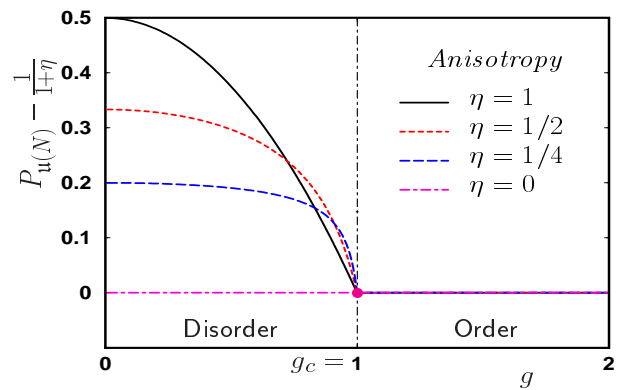}
\caption{Purity $P_{\fu(N)}$ for the BCS state as a function
of $g$. $P_{\fu(N)}$ scales with an exponent $\nu=1$ near $g_c$. 
Thus the correlation length diverges as $(g_c-g)^{-\nu}$ 
(Ising universality class). }
\label{QFT}
\end{figure}

{\bf Conclusion.$-$} We have introduced a generalization of
entanglement which goes beyond the standard subsystem-based approach
by considering entanglement as a quantum feature of states with
respect to any physically relevant, distinguished subspace of
observables.  \shortqph{These subspaces arise naturally from the algebraic
languages~\cite{Batista} used to describe quantum systems.}  In addition
to tying together the theory of entanglement and the theory of
coherent states, our results carry the potential for a number of
conceptual and practical advances. From a condensed-matter
perspective, GE might naturally provide measures of correlation
strength useful for establishing, for example, whether interactions
within a given quasiparticle description are sufficiently weak for a
mean-field theory to be meaningful.  Conversely, one might use a
typology of GE to better understand situations where mean-field theory
is not easily applied. For QIP, our formalism can give additional
insight into standard entanglement theory. It suggests novel
entanglement measures for the multipartite case.  By scaling system
sizes, asymptotic measures can be obtained to help investigate
information-theoretic or thermodynamic limits, with possible uses in
renormalization group analyses.
\shortqph{Finally, because the occurrence of a superselected structure 
in a quantum system provides an important avenue for effectively restricting 
the set of physically accessible operations, and can be formally associated 
to the reducible action of an appropriate operator set, our framework may 
provide further insight on the issue of entanglement in the presence
of superselection rules as recently addressed in~\cite{Verstraete2003,
Bartlett2003}. }

\acknowledgments
We thank C.D. Batista, J.E. Gubernatis, L. Gurvits and J. Preskill for 
discussions, and the DOE and NSA for support.

\bibliography{cs}

\end{document}

%% file: ekqc_a.tex
\usepackage{calc,ifthen}

\newlength{\elimdepthdim}
\newlength{\elimheightdim}
\newlength{\elimwidthdim}

\newlength{\strutdepthdim}
\newlength{\strutheightdim}
\newlength{\strutwidthdim}

\def\one{{\mathchoice {\rm 1\mskip-4mu l} {\rm 1\mskip-4mu l} {\rm
1\mskip-4.5mu l} {\rm 1\mskip-5mu l}}}

\newcommand{\ket}[1]{\qvbar{#1}\qrangle}
\newcommand{\bra}[1]{\qlangle{#1}\qvbar}

\newcommand{\lie}[1]{\mathfrak{#1}}

\newcounter{herefignum}

%% file: shortcohqph.bbl
\begin{thebibliography}{25}
\bibitem[{com({\natexlab{a}})}]{comment1}
\bibinfo{note}{A modified Lie bracket $i[\,,\,]$ is understood throughout.}
\bibitem[{com({\natexlab{b}})}]{comment2}
\bibinfo{note}{$\fh_c$ is constructed by taking the complex linear span of a
  basis for $\fh$. $\fh_c \oplus \openone$ guarantees inclusion of the
  $\openone$.}
\bibitem[{com({\natexlab{c}})}]{comment3}
\bibinfo{note}{Notice that the relevant representation is reducible in this
  case. However, the purity definition still applies.}
\end{thebibliography}
